\documentclass[aps,prl,twocolumn,superscriptaddress,showpacs,floatfix]{revtex4}

\usepackage{graphicx}

\begin{document}


\title{A Phase Diagram for Quantum Hall Bilayers with Strong Inter-layer Correlation}

\author{Biswajit Karmakar}
\email{b.karmakar@sns.it}
\affiliation{NEST INFM-CNR and Scuola Normale Superiore, Pisa 56126, Italy}
\author{Vittorio Pellegrini}
\affiliation{NEST INFM-CNR and Scuola Normale Superiore, Pisa 56126, Italy}
\author{Aron Pinczuk}
\affiliation{Depts of Appl. Phys \& Appl. Math. and of Physics, Columbia University, New York 10027, USA}
\author{Loren N. Pfeiffer}
\author{Ken W. West}
\affiliation{Bell Laboratories, Alcatel-Lucent, Murray Hill, New Jersey 07974, USA\footnote{Present address: Department of Electrical Engineering,
Princeton University, Princeton, NJ, USA}}




\date{\today}

\begin{abstract}

Inter-layer excitonic coherence in a quantum Hall bilayer with
negligible tunneling is monitored by measurements of low-lying spin
excitations. At $\nu_T =1$ new quasiparticle excitations
are observed above a transition temperature revealing a competing metallic phase. For
magnetic fields above an onset Zeeman energy this metallic phase has
full spin polarization. A phase diagram in the parameter space of temperature
and Zeeman energy reveals that the transition temperature increases
at higher fields. This unexpected result suggests intriguing impacts
of spin polarization in the highly correlated phases.
\end{abstract}


\maketitle Two-dimensional electron systems in double quantum wells
in the limit of vanishingly-small tunneling gap ($\Delta_{SAS}
\rightarrow 0$) display inter-layer coherence 
in the quantum Hall (QH)
states at total filling factor $\nu_T =1$ (just one Landau level
fully occupied) \cite{macbook,spielman}. This inter-layer correlated
quantum state underlying the QH effect is now
understood as an easy-plane pseudospin ferromagnet (up/down
pseudospin labels the electron occupation in the two layers).
Equivalently, the state can be viewed as a Bose-Einstein condensate
of inter-layer excitons by making a particle-hole transformation
(see Fig.1A) \cite{nature}. The strongly correlated excitonic phase
is incompressible with gapped low-lying collective excitations that
are linked to superfluid-like signature for the excitonic transport
such as seen in counterflow current experiments \cite{moon, kellog,
kellog1, tutuc, von, nature}.  In counterflow the onset of
inter-layer coherence occurs when the parameter $d/l_B$ is made
smaller than $\approx $ 1.9 ($d$ is the interlayer distance and $l_B$
the magnetic length)
\cite{spielman,spielman1,kellog,kellog1,tutuc,von}. Further, recent
magneto-transport experiments have stressed the role of spin
\cite{giudici} and focussed on the finite temperature properties of
the broken-symmetry excitonic phase \cite{champagne}. Recent theory
of inter-layer coherence in bilayer graphene suggests that such
superfluid-like behavior might be observable up to room temperature
\cite{macdonald02,macgra}.
\par
Theoretical formulations have raised the possibility that the
excitonic fluid competes with compressible (metallic) states of
composite-fermions (CF) \cite{simon,moller}. In these interpretations a CF phase,
metallic or BCS-like, occurs in the limit of large $d/l_B$ when the
system is regarded as two single layers each of them at $\nu =1/2$ .
The emergence of a Fermi sea of CFs has key manifestations in
excitations above the quantum ground state. In the low-lying sector
of spin modes there is a continuum of spin-flip excitations from
transitions between spin-up and spin down states across the CF Fermi
energy \cite{irene}. The quasiparticle transitions in this scenario
are shown in Fig.1B for a bilayer structure. The presence of such
spin excitations in spectra of bilayers are signatures of CF phases.
Studies of this class of low-lying spin modes thus offer direct
insights on the interplay between excitonic and CF phases in
bilayers with strong inter-layer correlation.
\par
We report here an experimental study of low-lying spin excitations
in QH bilayers at $\nu_T =1$ and with $\Delta_{SAS} \rightarrow 0$.
The modes are observed by resonant inelastic light scattering. The
experiments probe the regime where inter-layer coherence and
excitonic superfluidity have been reported. At the higher
temperatures the spectra show low-lying spin-flip modes below the Zeeman
energy at $E_Z$ similar to the CF excitations reported in single
electron layers at $\nu\rightarrow1/2$ \cite{irene}. We surmise that at the
higher temperatures the two layers are largely decoupled and that
each layer is in a metallic state with characteristic low-lying
spin-flip excitations such as to those of a CF Fermi sea in a single
electron layer at $\nu\rightarrow1/2$. At lowest temperatures and
$\nu_T =1$ there is near-absence of low-lying spin-flip modes. This
demonstrates the emergence of inter-layer coherence of an excitonic
phase. Slight filling factor deviations from $\nu_T=1$ restore the
signatures of low-lying spin-flip modes.
\par
We have carefully monitored the evolution of low-lying spin
excitations as a function of temperature and Zeeman energy. With
these results we construct a finite-temperature phase diagram for
the interplay between excitonic and metallic phases of bilayers at
$\nu _T =1$. We find that full spin polarization of the metallic
phase with CF signatures occurs at an onset Zeeman energy $E_
Z^*\simeq 0.12$ meV that is reached at fields 
$B_T^* \approx 5$T.
Surprisingly, we find that the transition temperature for the
transformation to the CF phase significantly increases for Zeeman
energies $E_Z\geq E_Z^*$. The unexpected behavior of the critical
temperature could be linked to the emergence of a novel correlated 
phase at intermediate temperatures. Our results therefore highlight a rich phase diagram
with multiple phases dictated by competition between quantum and
thermal effects.
\par
Measurements were performed on the sample mounted on a mechanical
rotator in a dilution refrigerator with base temperature of 50 mK 
under light illumination. The sample is a nominally symmetric
modulation-doped
$\textnormal{Al}_{0.1}\textnormal{Ga}_{0.9}\textnormal{As}/\textnormal{GaAs}$
double quantum well structure with $AlAs$ barrier in between the wells, 
having well width of 18nm and barrier width of 7nm. The large
barrier ensures that the tunneling gap is vanishingly small
($\Delta_{SAS} \rightarrow 0$). The total electron density is $n_T
\sim 6.9 \times 10^{10}$ $\textnormal{cm}^{-2}$ and electron
mobility above $10^6$ $\textnormal{cm}^2/\textnormal{Vs}$. A
perpendicular magnetic field of 2.85T was applied to bring the
electron bilayer in the quantum Hall (QH) state corresponding to
$\nu_T =1$. The sample has $d/l_B = 1.65$. Therefore electrons at
sufficiently low temperatures are in the correlated phase where
exciton condensation and counterflow superfluidity take place. Spin
excitations were measured by resonant inelastic light scattering in
a backscattering configuration at different tilt angles $\theta $
between the magnetic field direction and the plane of the sample
(see left part of Fig.2). To this end a single-mode tunable
Ti-Sapphire laser at around 810 nm and a triple-grating spectrometer
equipped with a CCD detector were used. Laser power densities were
kept at $\sim~ 10^{-4}$ $\textnormal{W/cm}^2$ to avoid electron
heating effects and a crossed polarization configuration with
perpendicular incident and scattered photon polarizations was
exploited to have access to spin modes \cite{luin05} (see
supplementary materials).
\par
\par
\begin{figure}
\begin{center}
\includegraphics*[width=8cm]{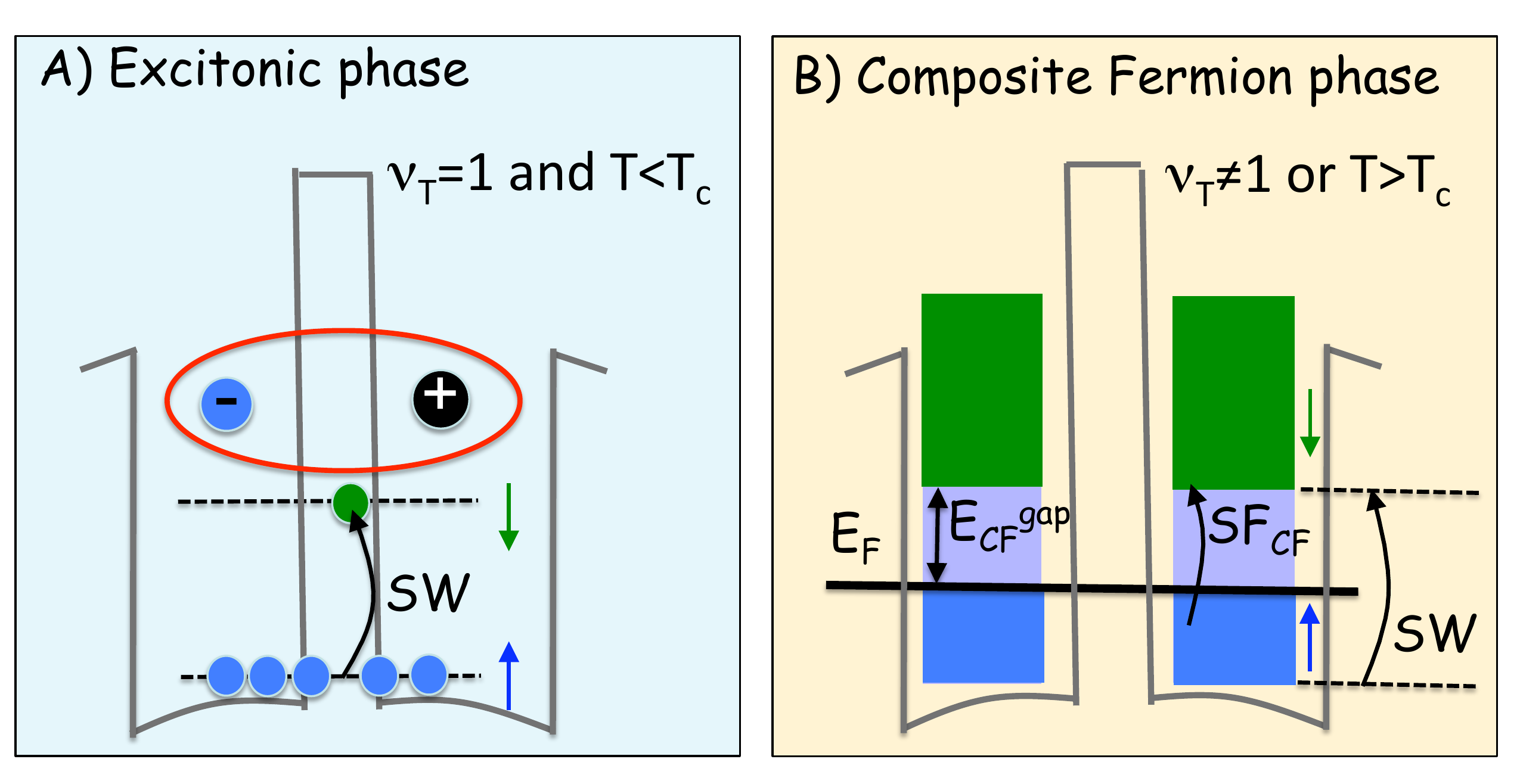}
\end{center}
\caption{A: Particle-hole transformation in bilayers at total filling factor $\nu _T =1$ allows to describe the population of one of the two
layers in terms of holes. The resulting ground state consists of a condensate of interlayer excitons and a conventional quantum Hall fluid formed by
the excess charge that fully occupy the spin-up Landau level and supports a well-defined spin-wave (SW) mode across the Zeeman gap.
B: Schematic diagram of composite-fermion (CF) levels and spin excitations  of the bilayer CF metal.
$E_F$ is the CF Fermi energy. In the CF phase, a spin flip ($\textnormal{SF}_{\textnormal{\scriptsize{CF}}}$) continuum of excitations across the Fermi level extends from the Zeeman gap down to an energy value $E_{CF}^{gap}$ determined by the relative position of the Fermi level within the spin-up and spin-down CF states. The drawing shown in the figure refers to the case of a spin-polarized CF metallic state.}
\end{figure}
\par
Figure 1 describes the energy level structure and electronic
spin-fip excitations of the excitonic QH phase (A) and of the Fermi
seas of CFs (B) that occur when inter-layer coherence is lost. We
recall that a CF quasiparticle can be viewed as an electron with an
even-integer number of magnetic flux quanta attached to it
\cite{jain}. As a consequence, the CFs sense an effective magnetic
field that is zero when the electron filling factor is $\nu = 1/2$.
In this limit the spacing between the CF Landau levels vanishes and
the CFs coalesce into a Fermi sea \cite{halperin} as shown in Fig.1
right panel in the case of double layers with 
$\nu _T$ = ($\nu$=1/2) + ($\nu$=1/2) = 1.
\par
In the excitation spectra, the formation of a Fermi sea of CFs
manifests into a continuum of spin-flip modes between spin-up and
spin down states across the CF Fermi energy (see the diagram in
Fig.1B) \cite{irene, karmakar, karmakar1}. Such excitations, that
are accessible by inelastic light scattering methods, represent the
hallmark of the CF phase in bilayers. As shown below these
collective modes offer direct probes of the manifestation of phases
with strong inter-layer excitonic correlation. The collective 
electronic spin-flip excitation in the excitonic phase, instead, can
be understood by considering the excess spin-up electrons that
remain after the particle-hole transformation. These electrons fill
exactly one Landau level and support a well-defined long wavelength
SW mode at the Zeeman energy as shown in Fig.1A.
\par
Figure 2 reports representative low-lying spin excitations measured
as function of small changes in filling factor near $\nu_T =1$. The
spectra reveal a well-defined SW mode with Lorentzian lineshape and
the continuum of CF spin excitations with a pronounced minimum when
the bilayer is in the QH state at $\nu_T =1$. This significant
result indicates a trend for the compressible CF metal to transform
into a new phase with inter-layer coherence when the total filling
factor approaches $\nu_T =1$. Observation of residual 
($\textnormal{SF}_{\textnormal{\scriptsize{CF}}}$) modes seen at $\nu
_T =1$ at the lowest temperatures (see Fig.3 also) is consistent with co-existence of
compressible-incompressible puddles due to disorder in
agreement with previous experimental and theoretical analysis
\cite{kellogg,luin06,stern,fertig}. No spin-flip mode is seen above
the SW. This mode is indeed observed in samples with finite values
of $\Delta _{SAS}$ \cite{karmakar1, luin05}. Its absence supports
the conclusion of negligible interlayer tunneling in the sample
studied here.
\par
\par
\begin{figure}
\begin{center}
\includegraphics*[width=9.5cm]{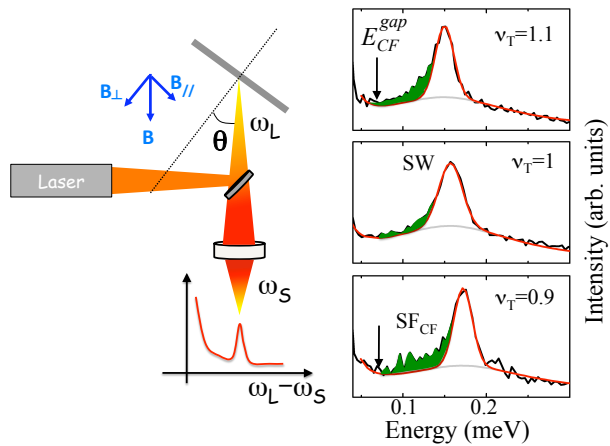}
\end{center}
\caption{Left panel: Geometry of the experimental set-up for
inelastic light scattering. $\omega _L$ and $\omega _S$ refer to
incident and scattered photon frequencies (see supplementary
materials for a description of the inelastic light scattering
mechanism). Representative spin excitation spectra at $T=50$ mK and
three values of filling factor $\nu _T$ at a tilt angle of $\theta = 67.5^o$.
A lorentzian fit to the spin wave (SW) is shown (red line) to
highlight the impact of the continuum of spin-flip excitations
$\textnormal{SF}_{\textnormal{\scriptsize{CF}}}$ (shaded in green).
The gray line is the background due to the magneto-luminescence.}
\end{figure}
\par
The observation of $\textnormal{SF}_{\textnormal{\scriptsize{CF}}}$
excitations demonstrates the emergence of a CF metallic phase and
illustrates how the inter-layer coherence of the QH fluid is lost by
varying the filling factor.  As inter-layer coherence disappears the
physics becomes dominated by the intra-layer correlations that build
up the CF quasiparticles. Particularly revealing of the competition
between the inter-layer excitons and the CFs is the behavior as a
function of temperature shown in Fig.3. The integrated intensity of
the spin continuum remains small and constant up to a critical value
of temperature $T_c$ and then it increases markedly. This anomalous
behavior also seen in magneto-transport characteristics of the
bilayer QH state \cite{lay,champagne} suggests a finite-temperature
phase transformation of the coherent QH state to the CF metal at
$T_c$.  Weather the transition exploits a Kosterlitz-Thouless
mechanisms as expected for the broken-symmetry bilayer state remains
an open issue that deserves additional experimental investigation.
\par
\begin{figure}
\begin{center}
\includegraphics*[width=8cm]{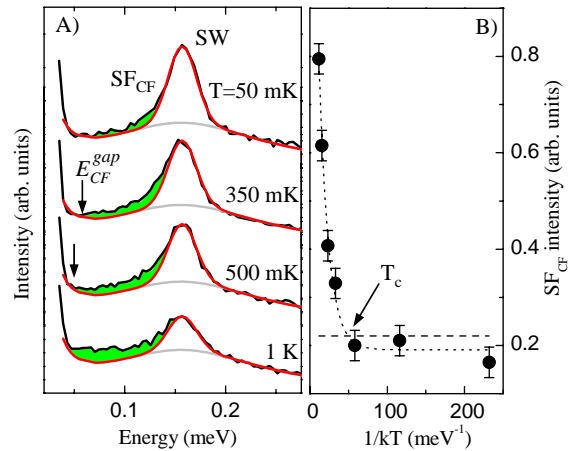}
\end{center}
\caption{A) Representative spin excitation spectra at $\nu _T =1$
and different temperatures. Red line is a Lorentzian fitting to the
SW. The continuum of spin-flip excitations
$\textnormal{SF}_{\textnormal{\scriptsize{CF}}}$ is shaded in green.
The gray line is the background due to the magneto-luminescence.
B) Evolution of the integrated intensity of
$\textnormal{SF}_{\textnormal{\scriptsize{CF}}}$ as a function of
the inverse of temperature. The dashed line represents the estimated
critical temperature $T_c$ signaling the transition to the
composite-fermion phase.}
\end{figure}
It can be noted that the lowest energy value $E_{CF}^{gap}$ of the
$\textnormal{SF}_{\textnormal{\scriptsize{CF}}}$  continuum remains
finite at the lowest temperature in the data displayed in Figs. 2
and 3, measured at the relatively large angle $\theta = 67.5^o$
($B_T = 7.46$T). This shows that the CF metal is fully
spin-polarized. The lowest-energy value of the spin-flip continuum,
however, lowers to values below the experimental resolution ($30 \mu
$eV) as temperature increases (see the lowest spectrum in Fig.3 at
$T = 1$K). The same happens as the angle decreases (data not shown). The
collapse of $E_{CF}^{gap}$ signals a spin transition to a CF state
with partial spin polarization. Further evidence of the spin
transition of the CF metal as a function of angle is also found in the behavior of the SW
intensity versus temperature (see supplementary materials).

\begin{figure}
\begin{center}
\includegraphics[width=10cm]{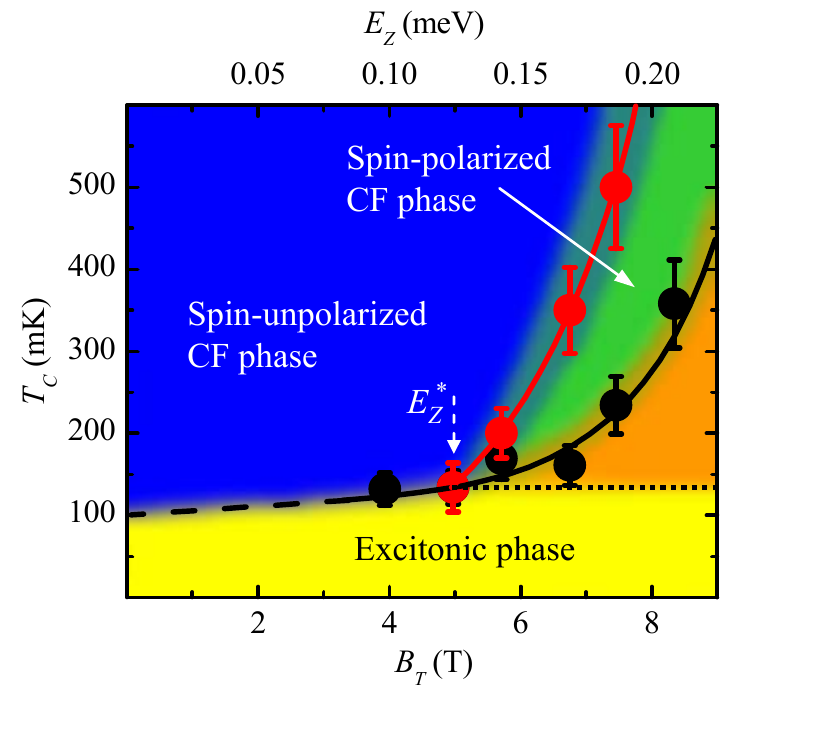}
\end{center}
\caption{Phase diagram for the competition between the
incompressible inter-layer coherent state and the compressible
bilayer composite-fermion metal (CF) at $\nu _T =1$. The transition
temperature is given as function of the total magnetic field $B_T$
and Zeeman energy $E_Z$. Green(blue) regions indicate
spin-unpolarized (spin-polarized) CF metals. The red points define
the temperature at which the spin-flip continuum extends down to
below $30 \mu$eV marking the loss of spin-polarization in the
spin-polarized CF metal. The orange region at large total fields
defines a possible intermediate phase. The dotted line is the
hypothetical boundary between the inter-layer excitonic and
intermediate phases. $E^{*}_{Z}$ represents the critical value of
the Zeeman energy that marks the transition from the
spin-unpolarized to spin-polarized CF metal.}
\end{figure}
\par
The phase diagram shown in Fig. 4 is obtained by reporting the
measured values $T_c$ at different values of the tilt angle while
keeping the filling factor at $\nu _T =1$.  This phase diagram
highlights an intriguing competition between a low-temperature
region where inter-layer correlation dominates with the
high-temperature CF phase. It also shows that a spin transition to a
fully spin-polarized CF state occurs at a critical magnetic field of
$B^{*}_{T} \approx 5$T.
\par
The increase of $T_c$ for magnetic fields above $B^{*}_{T} = 5$T is
unexpected. The reason is that no further change in the boundary
between the two phases is expected for 
$B_{T} \geq 5$T 
when also the CF state becomes fully spin-polarized (see the horizontal dotted
line in Fig.4) \cite{giudici,champagne}.  Additionally, at large
$\theta $ values, the in-plane magnetic field leads to the
compression of the electron wavefunction along the z-direction.
Although this orbital effect is expected to be rather small in our
quantum wells \cite{giudici}, it increases the intra-layer
interactions responsible for the formation of CF quasiparticles
\cite{gee} and should favor the CF phase. Again, this scenario is in
contrast to the behavior shown in Fig.4.
\par
We conjecture that the increase in $T_c$ for 
$B_{T}\geq B^{*}_{T} = 5$T 
could be evidence for the formation of an intermediate
non-CF phase. Unlike the coherent phase of inter-layer excitons that
occurs at the lowest temperatures below the dotted line in Fig.4, this intermediate
phase could have residual but non negligible inter-layer correlation. It
is tempting to associate this additional phase to the formation of a
new ground state due to coupling between CFs in the two layers.
Recent predictions have indeed shown that a QH state due to CF
pairing could lead to an intermediate phase between the coherent
excitonic and CF states with a BCS order parameter
\cite{moller,bonesteel}. As the inter-layer correlation further
increases at lower temperatures this intermediate phase develops
eventually into the excitonic state. Further experiments are needed
in this regime to test this interpretation. We would like to remark, however,
that in transport experiments at large tilt angles, the disappearing
of the excitonic phase is monitored by the collapse of the QH effect.
This observation seems to suggest 
that the intermediate phase is compressible.

{\bf Supplementary materials}

{\bf Mechanism for resonant inelastic light scattering}
Within time-dependent perturbation theory the resonant inelastic light scattering cross section includes two virtual intermediate
valence-to-conduction band optical transitions. The resonance enhancement occurs when the incoming and/or scattered photon overlaps in
energy a real interband transition across the GaAs band gap. Conservation of energy implies that $\hbar \omega_I - \hbar \omega_S = \pm \hbar \omega ({\bf q})$ where $\hbar \omega _{I,S}$ represents the energy of the incoming or scattered photon and $\hbar \omega ({\bf q})$ the energy of the collective mode with in-plane wavevector ${\bf q}$. When the two-dimensional electron gas has full translational invariance there is conservation of in-plane wavevector in the light scattering process. In this case the transferred component of the wavevector ${\bf q}$ is related to the photon momentum in the crystal and with infrared source is small and typically in the range $q\cdot l_B \le 0.2$ where $l_B$ is the magnetic length \cite{pss}.  
\par
Inelastic light scattering by excitations with reversal of the spin such as the spin-wave (SW) exploits p-like states at the top of the valence band such as light-hole states with mixed spin and orbital character due to spin-orbit coupling. In addition the incoming or scattered light must have a component of the polarization along the z-axis perpendicular to the QW plane which requires a tilted field geometry with a non-zero angle $\theta$. The cross section for spin excitations is proportional to ${\bf e}_I$ x ${\bf e}_S$ where ${\bf e}_{I(S)}$ are the incoming (scattered) photon polarizations \cite{yafet}.
\par
\begin{figure}
\begin{center}
\includegraphics[width=7cm]{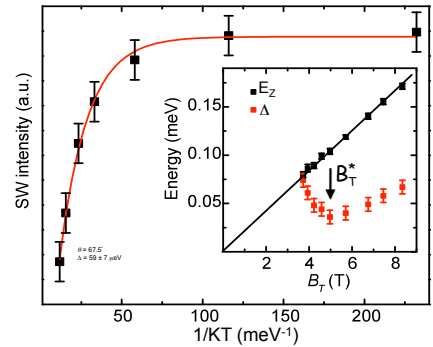}
\end{center}
\caption{Integrated intensity $I$ of the SW peak as a function of temperature. The red line represents the best-fit result with the function $I(T) = I_o (1-e^{\Delta /kT}$). The inset shows the behavior of $\Delta $ as a function of total magnetic field $B_T$. The peak energy of the SW peak that corresponds to the Zeeman energy $E_Z$ is also shown. The black line is a fit to the equation $E_Z = g\mu_B B_T$ where $\mu _B$ is the Bohr magneton and $g=-0.4$ the gyromagnetic factor.}
\end{figure}
\par
{\bf Activation energy of the spin wave}
Further evidence of the spin transition of the CF metal is also found in the behavior of the SW intensity versus temperature. The integrated intensity of the Lorentzian fit to the SW reported in Fig.1 as a function of the inverse of temperature for $\theta = 67.5^o$ remains constant up to $\approx 150 mK$ and then displays an activated behavior. The resulting activation gap $\Delta $ depends on $B_T$ as shown in the inset to Fig.1 with a minimum at 5T that signals a change in the spin properties of the CF phase. For comparison the values of $E_Z$, identified as the peak energy of the SW mode as predicted by the Larmour theorem, is also plotted in the inset to Fig.1. Further work is required to develop the microscopic mechanism underlying the behavior shown in Fig.1

\begin{acknowledgments}
This work was supported by the projects MIUR-FIRB No. RBIN04EY74 and
by the Nanoscale Science and Engineering Initiative of the National
Science Foundation under NSF Award Number CHE-0641523, A. P. is
supported by the National Science Foundation under Grants No.
DMR-0352738 and DMR-0803445, by the Department of Energy under Grant
No. DE-AIO2-04ER46133, and by a research grant from the W. M. Keck
Foundation. Useful discussions with K. Muraki and J.P. Eisenstein
are gratefully acknowledged.
\end{acknowledgments}


\end{document}